\documentclass[sigplan,screen]{acmart}
\copyrightyear{2022} 
\acmYear{2022} 
\setcopyright{acmcopyright}\acmConference[ESSE 2022]{2022 The 3rd European Symposium on Software Engineering}{October 27--29, 2022}{Rome, Italy}
\acmBooktitle{2022 The 3rd European Symposium on Software Engineering (ESSE 2022), October 27--29, 2022, Rome, Italy}
\acmPrice{15.00}
\acmDOI{10.1145/3571697.3571704}
\acmISBN{978-1-4503-9730-8/22/10}
\usepackage{array,multirow,graphicx}
\usepackage{pgfplots}
\usepackage[T1]{fontenc}
\usepackage{tikz}
\usepackage{array,booktabs}
\usepackage{makecell, pict2e}
\usetikzlibrary{patterns}
\pgfplotsset{width=16cm,compat=1.8}
\usepgfplotslibrary{groupplots}
\usepackage{algorithmic}
\usepackage{diagbox}
\usepackage{graphicx}
\usepackage{textcomp}
\usepackage{xcolor}
\usepackage{lipsum,tabularx}
\usepackage{booktabs,graphicx}
\usepackage{caption,subcaption}
\usepackage{hyperref}
\usepackage{rotating}
\usepackage{array,makecell,multirow}

\newcolumntype{P}[1]{>{\centering\arraybackslash}p{#1}}
\begin{document}

\title{Qualitative analysis of the relationship between design smells and software engineering challenges}


\author{Asif Imran}

\email{aimran@csusm.edu}
\orcid{0000-0002-1780-0296}

\affiliation{%
  \institution{California State University San Marcos}
  \streetaddress{333 S Twin Oaks Valley}
  \city{San Marcos}
  \state{California}
  \country{USA}
  \postcode{92096}
}

\author{Tevfik Kosar}

\email{tkosar@buffalo.edu}
\orcid{0000-0002-5600-6706}

\affiliation{%
  \institution{University at Buffalo}
  \streetaddress{Davis Hall}
  \city{Buffalo}
  \state{New York}
  \country{USA}
  \postcode{14260}
}


\begin{abstract}
Software design debt aims to elucidate the rectification attempts of the present design flaws and studies the influence of those to the cost and time of the software. Design smells are a key cause of incurring design debt. Although the impact of design smells on design debt have been predominantly considered in current literature, how design smells are caused due to not following software engineering best practices require more exploration. This research provides a tool which is used for design smell detection in Java software by analyzing large volume of source codes. More specifically, 409,539 Lines of Code (LoC) and 17,760 class files of open source Java software are analyzed here. Obtained results show desirable precision values ranging from 81.01\% to 93.43\%. Based on the output of the tool, a study is conducted to relate the cause of the detected design smells to two software engineering challenges namely \textit{"irregular team meetings"} and \textit{"scope creep"}. As a result, the gained information will provide insight to the software engineers to take necessary steps of design remediation actions.

\end{abstract}



\maketitle

\section{Introduction}
\label{introduction}

Software design engineering is an important activity which requires careful application of design guidelines. Design issues contribute to 64\% of software defects as a study by Jones et.al. \cite{softqual} highlighted. Hence, quality and maintainability of software are significantly affected by design problems. One of the important design issues is a software suffering from design smells. Design smells violate the architectural principles and negatively affect the design of the software \cite{suryanarayana2014refactoring}. A software which has large volume of design smells contributes to design debt which in-turn increases technical debt. In recent years, a greater emphasis to reduce design smells has been given by software companies, engineers and researchers. Despite the increased importance, both developing and established software suffer from design smells, which is undesirable and hampers the long term sustainability. Research efforts are needed to identify the issues in software engineering practices deployed by companies which results in the design smells. 

List of 25 design smells were provided by Suryanarayana et al. \cite{suryanarayana2014refactoring} which focused on all the fundamental design aspects. A design debt prioritization using portfolio matrix was proposed by Plosch et al. \cite{plosch2018design}. A model for detecting cyclic dependency and hub-like dependency smells using link prediction techniques was discussed by Diaz-Pace \cite{diaz2018towards}. A catalogue to list all architectural smell detection tools together with their operating platforms was provided by Azadi et al. \cite{azadi2019architectural}. However, the accuracy of the tool requires further improvement. Additionally, teamwork issues in real life software development that contributes to design smells were not analyzed.

Based on the above motivation, this paper contributes primarily to determining a tool for design smell detection followed by establishing a relationship between the occurrence of a design smell to software engineering challenges in real life software development. First, the paper proposes a tool for design smell detection based on pseudo-model generation using \textit{Abstract Syntax Trees (AST)}. It is used to analyze 409,539 Lines of Code (LoC) and 17,760 class files of open source Java software. Finally, precision and recall are calculated to identify the performance of the tool by 16 industry experts. Next, a survey is conducted with the same experts to identify a relationship between presence of the detected design smells to software engineering challenges like \textit{"irregular team meetings"} and \textit{"scope creep"}.

This paper focuses on nine design smells highlighted in Table \ref{designsmellstable} which were selected from existing literature \cite{suryanarayana2014refactoring}. We also obtained source codes of release versions for 11 software which formed the testbed for our experiments. The names of the software have not been presented due to confidentiality issues and they are addressed as $S_1..S_{11}$ respectively. Metadata about the software are highlighted in Table \ref{list}. A total of 4,020 design smells are detected in our study. Quantitative values of the frequency of occurrence for all the smells is provided in the results analysis section. The results are desirable as precision values ranging from \textit{81.01\%} to \textit{93.43\%} are obtained for the analyzed software. Next, we conduct a qualitative study to identify the cause of the smells based on two software engineering challenges namely \textit{"irregular team meetings"} and \textit{"scope creep"}. These two challenges are selected mainly due to their influence on quality software development \cite{aniche2019current}. We conduct a survey among the 16 software experts with an aim to determine this relationship. The findings show significant impact of the two issues on the occurrence of design smells. 

Based on the above information, the major contributions of this paper can be stated as follows:-
\begin{itemize}
    \item Provide a tool for design smell detection for Java software and analyze its performance.
    \item Establish relationship between key software engineering challenges and the occurrence of design smells.
\end{itemize}

Rest of the paper proceeds as follows.  Section \ref{empirical} illustrates the research questions, identifies the selected software systems for this study and describes the implementation of the tool for design smell detection and recording. Section \ref{analysis} provides the obtained results and analyses those. Section \ref{survey} describes the conducted survey to establish relationship between software engineering challenges and occurrence of design smells. Section \ref{threat} highlights the threats to validity and credibility, Section \ref{relatedwork} discusses the related work, and Section \ref{conclusion} concludes the paper and discusses future research directions.

\begin{table*}
    \centering
    \scalebox{1}{
    \begin{tabular}{|p{1.0\textwidth}|}
        \hline \textbf{Unutilized Abstraction}: This design smell occurs when an abstraction is declared, however its implementation is missing in the code. For example, a class \textit{ABC} may be declared by the software developer, however if it is not used anywhere in the code, then it is a \textit{Unutilized Abstraction} design smell.\\
        \hline\hline
        \textbf{Insufficient Modularization}: When an abstraction is large and complicated, providing a scope of further modularization, it is said to suffer from this design smell. This smell occurs when the class is large in size, multiple class definitions are present in a file or the complexity of the class is high.\\
        \hline\hline
        \textbf{Broken Hierarchy}: This occurs when the IS-A linkage between the supertype and subtype is absent. This interferes with the substitution capability among the two classes.\\
        \hline\hline
        \textbf{Deficient Encapsulation}: When an abstraction given more accessibility to the users than it is required, it threatens the security of the software. The presence of this smell is called \textit{Deficient Encapsulation}.\\
        \hline\hline
        \textbf{Cyclic Dependent Modularization:} This smell occurs when the software violates the acyclic modularization technique. It occurs when one abstraction is cyclically dependent on another. If this design smell persists, then a change in one abstraction can have a ripple effect on other abstractions in the software. If the cyclic relationship is complicated in a  large software, it is a challenge to detect that smell.\\
        \hline\hline
        \textbf{Unnecessary Abstraction:} When a software has multiple layers of abstraction and many of the intermediate layers are not required, the software is said to suffer from \textit{Unnecessary Abstractions}.\\
        \hline\hline
        \textbf{Wide Hierarchy}: This smell is prevalent when an inheritance tree has a wide breadth and the intermediate types are missing. Our experiments show that it may appear in smaller volume in the software.\\
        \hline\hline
        \textbf{Missing Hierarchy}: It occurs when a class has limited functionality within an operation. It may also happen when an operation is covered to a class. It can have a significant influence on the object oriented design, hence it is considered a design smell.\\
        \hline\hline
        \textbf{Multifaceted Abstraction}: When an abstraction is responsible for multiple functionalities, managing it may become troublesome, resulting in this design smell . Also, many functionalities might be affected if an error occurs in the abstraction.\\
        \hline
    \end{tabular}}
    \caption{Description of the selected design smells}
    \label{designsmellstable}
\end{table*}

\section{Empirical study setup}
\label{empirical}
This qualitative study focuses on detecting design smells and identifying their causes based on feedback from industry experts. The following subsections identify the research questions and data collection strategy.
\subsection{Research Questions}
The following research questions are addressed in this paper.\\ 

\textbf{RQ1: How to detect design smells in Java source codes?}

This paper provides a tool to detect design related smells. The answer to this research questions will help software engineers detect design smells in the code, hence saving critical design debt and time.\\

\textbf{RQ2: To what extent does real life software engineering challenges impact the presence of design smells?}

The software team faces critical challenges like \textit{"irregular team meetings"} and \textit{"scope creep"} which impacts the design of the software. Such issues may also result in the occurrence of design smells in the final software product. We aim to study this relationship here.
\subsection{Selection of Software Systems}

\begin{figure}
    \centering
    \includegraphics[width=5.8cm, height=6cm]{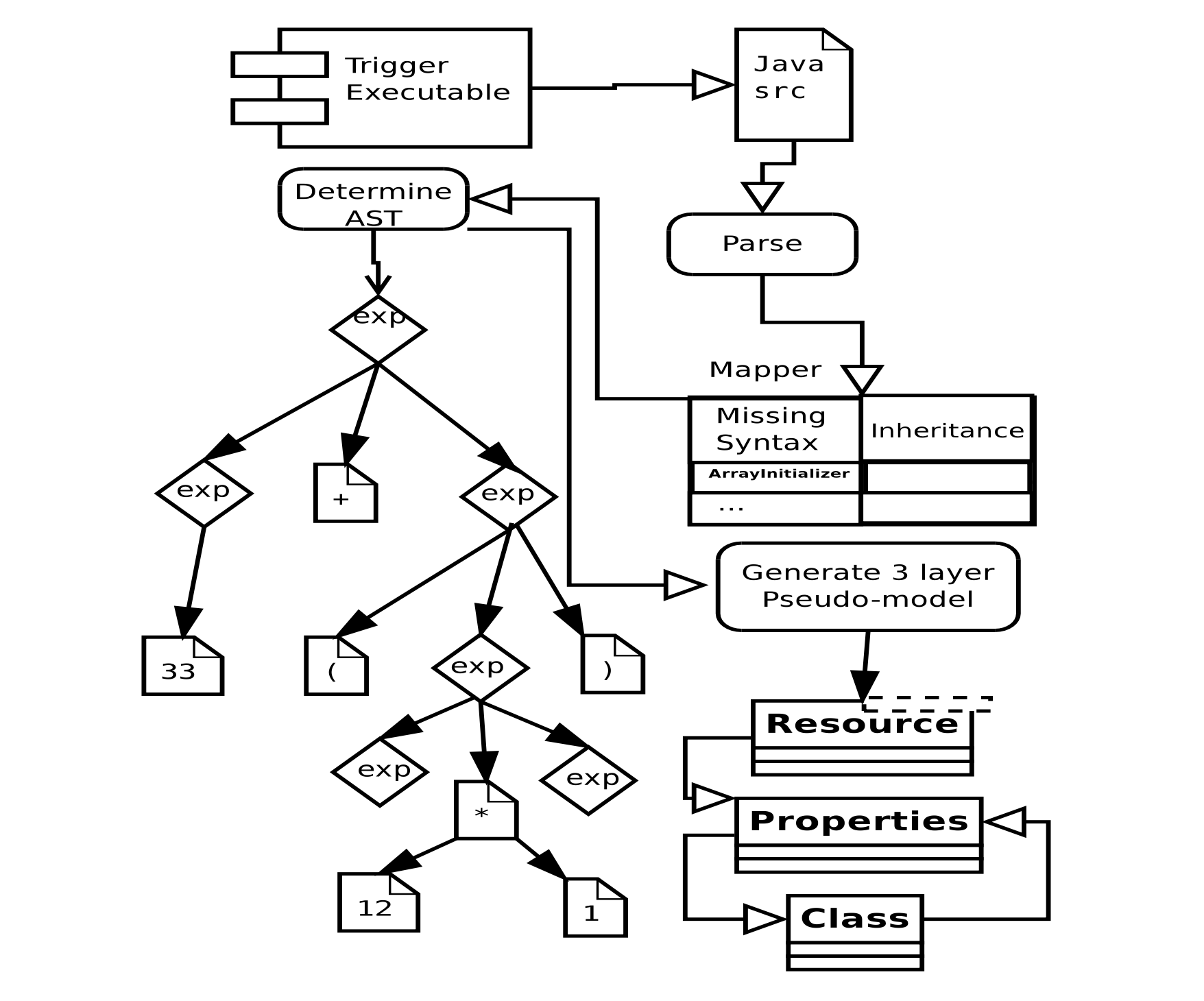}
    \caption{Representation for generating AST and pseudomodel for design smell analysis}
    \label{fig:framework}
\end{figure}

Table \ref{list} summarizes the list of eleven selected systems. For each of the systems, we highlight the life-time, $KLoc$, number of commits. Next, we tested those software to detect the design smells using the tool described in the following subsection.

\subsection{Implementing Smell Detection Tool} The tool described here is motivated by the \textit{Designite} tool \cite{7809461}. However, \textit{Designite} is designed for detection of smells for C\# code whereas our tool is modified for detecting design smells in Java code. Figure \ref{fig:framework} identifies the process of \textit{AST} and pseudo-model generation for obtaining the design smells. It is stated that 33.27\% of the software in the world are coded using Java. Presence of design smells in Java code significantly challenges the software engineer during incorporation of new updates. Also design smells possesses significant issues during final deployment of the software. Therefore, we are motivated to provide a tool to Java developers which can be used to detect design smells in their software, thus give them information related to the design smells present in their code. 

Firstly, Java source codes are taken as input which is then used to form an \textit{Abstract Syntax Tree (AST)}. The tokenized source text is read, to convert it to a tree. The first phase includes lexical analysis followed by syntax analysis. To form the \textit{AST, Janett} \cite{janett} has been used which converts constructs and calls to Java libraries to C\# format. Hence the \textit{AST} can be formed using \textit{NRefactory} \cite{grunwald2012nrefactory} after parsing the Java code with \textit{Janett}. Special Java constructs like $Abstract classes$, \textit{ArrayInitializer}, etc which are not present in C\# are converted using a mapper. \textit{IKVM} library is used to translate syntax and constructs, using inheritance whenever is necessary. Once the \textit{AST} is ready, it is read via a script which translates it to a simple pseudo model.

\begin{table}
\centering
 \begin{tabular}{| c | c | c | c |} 
 \hline
 System & Branch History &KLoC & Commits \\ [0.3ex] 
 \hline
\scriptsize $S_1$ &  \scriptsize 02/14-12/21 &\scriptsize 29.051 &\scriptsize 1871 \\ 
 \scriptsize $S_2$ &\scriptsize 09/14-03/21 &\scriptsize 64.448 &\scriptsize 88 \\
 \scriptsize $S_3$ &\scriptsize 05/14-02/18 &\scriptsize 23.002 &\scriptsize 1644 \\
 \scriptsize $S_4$ &\scriptsize 08/09-03/19 &\scriptsize 24.267 &\scriptsize 385 \\
 \scriptsize $S_5$ &\scriptsize 11/20-05/22 &\scriptsize 17.201 &\scriptsize 412 \\
 \scriptsize $S_6$ &\scriptsize 07/10-05/19 &\scriptsize 19.391 &\scriptsize 352 \\ 
 \scriptsize $S_7$ &\scriptsize 03/12-04/19 &\scriptsize 141.038 &\scriptsize 5531\\
 \scriptsize $S_8$ &\scriptsize 09/06-03/19 &\scriptsize 27.271 &\scriptsize 868 \\
\scriptsize $S_9$ &\scriptsize 08/05-03/19 &\scriptsize 24.698 &\scriptsize 15052\\
\scriptsize $S_{10}$ &\scriptsize 07/02-03/19 &\scriptsize 49.534 &\scriptsize 5446\\
\scriptsize $S_{11}$  &\scriptsize 10/19-11/21 &\scriptsize 17.331 &\scriptsize 788 \\
 \hline
\end{tabular}
\caption{List of selected systems for the study}
\label{list}
\end{table}

\begin{figure}
    \centering
    \includegraphics[width=7cm, height=3cm]{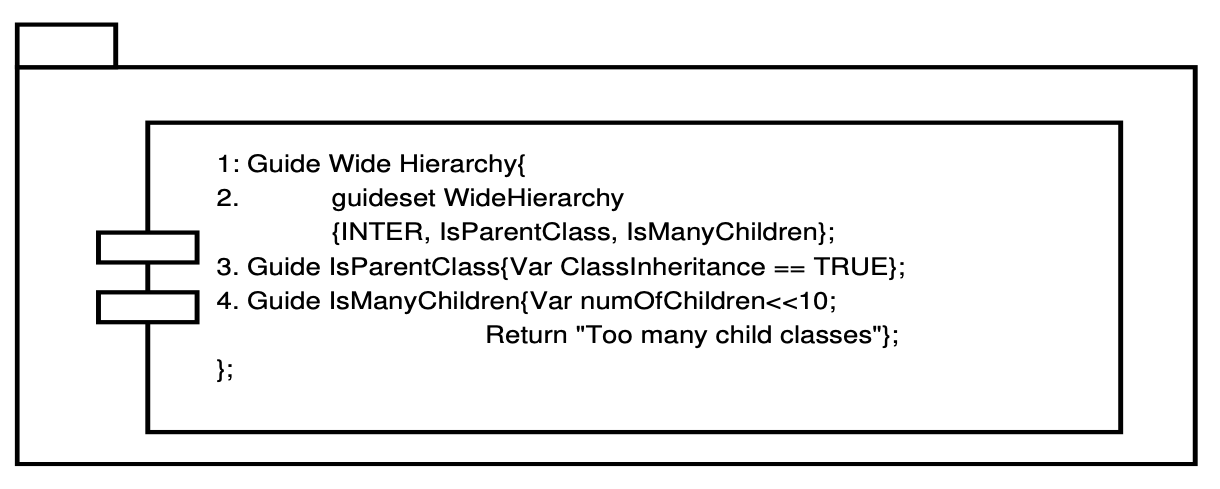}
    \caption{Rule card for detecting wide hierarchy design smell}
    \label{wideheirarchy}
\end{figure}

After following the stated steps we can have a simple version of pseudo-model from the $AST$ consisting of 4 layers. The lowest layer consists of the data elements and objects in Java code. The second layer is a description of the elements found in the lowest layer. The third layer consists of the data types and namespaces of the objects. Using the information captured in the pseudo-model, design smells in the Java code are detected with pre-specified rule cards \cite{8919054}. Example of a rule card for detecting wide hierarchy design smell is provided in Figure \ref{wideheirarchy}. Finally, those smells are written in a provenance file which can be used for analysis.

To use the system, users will require a Linux platform with terminal access. The \textit{CLI} will enable running the tool to obtain frequency of specific design smells for software. Firstly, the executable version of the tool needs to be triggered. This is followed by passing the path of the Java source files of the target software. Next, the parameter and threshold values are passed. Afterwards, the design smells can be detected. Once detected, the tool writes the output of detection to a $provenance.log$ file which can be later further analyzed considering it as a structured data-set of smells to obtain frequency of design smell prevalent in the software. 
\renewcommand{\arraystretch}{0.60}
\setlength{\tabcolsep}{1pt}
\newcommand\Tstrut{\rule{0pt}{2.6ex}}       
\newcommand\Bstrut{\rule[-0.9ex]{0pt}{0pt}} 
\newcommand{\TBstrut}{\Tstrut\Bstrut} 
\begin{table*}
\centering
\scalebox{1.4}{
\scriptsize
\begin{tabular}{|p{3cm}|p{0.6cm}|p{0.6cm}|p{0.6cm}|p{0.6cm}|p{0.6cm}|p{0.6cm}|p{0.6cm}|p{0.6cm}|p{0.6cm}|p{0.6cm}|p{0.6cm}|}
\cline{2-12}
\hline
{Design smells} &
 {$S_3$} & {$S_6$} & {$S_4$} & {$S_5$} & {$S_1$} & {$S_2$} & {$S_7$} & {$S_9$} & {$S_{10}$} & {$S_8$} & {$S_{11}$} \\ \hline
        Unutilized Abstraction & 164 & 69 & 96 & 56 & 99 & 341 & 884 & 104 & 114 & 56 & 89 \\ \hline
        Insufficient Modularization    & 7 & 9 & 28 & 3 & 22 & 38 & 189 & 62 & 48 & 41 & 43 \\ \hline
        Broken Hierarchy & 49 & 31 & 24 & 16 & 21 & 41 & 47 & 6 & 4 & 0 & 0 \\ \hline
        Deficient Encapsulation     & 39 & 33 & 49 & 19 & 24 & 62 & 18 & 21 & 29 & 16 & 27 \\ \hline
        Cyclic-Dependent Modularization   & 23 & 0 & 17 & 2 & 5 & 102 & 252 & 34 & 30  & 27  & 33 \\ \hline
        Unnecessary Abstraction   & 4 & 3 & 5 & 8 & 0 & 26 & 76 & 15 & 11  & 14  & 12 \\ \hline
        Multifaceted Abstraction   & 3 & 1 & 3 & 1 & 2 & 1 & 18 & 20 & 13  & 11  & 17 \\ \hline
        Wide Hierarchy   & 3 & 3 & 7 & 2 & 6 & 4 & 3 & 1 & 0  & 0  & 0 \\ \hline
        Missing Hierarchy   & 0 & 0 & 1 & 0 & 1 & 3 & 0 & 2 & 0  & 0  & 0 \\
  \hline
    \end{tabular}
    }
\caption{List of considered design smells}
\label{fig:top20list3}
\end{table*}
\begin{figure*}
    \centering
    \includegraphics[width=15cm, height=12cm]{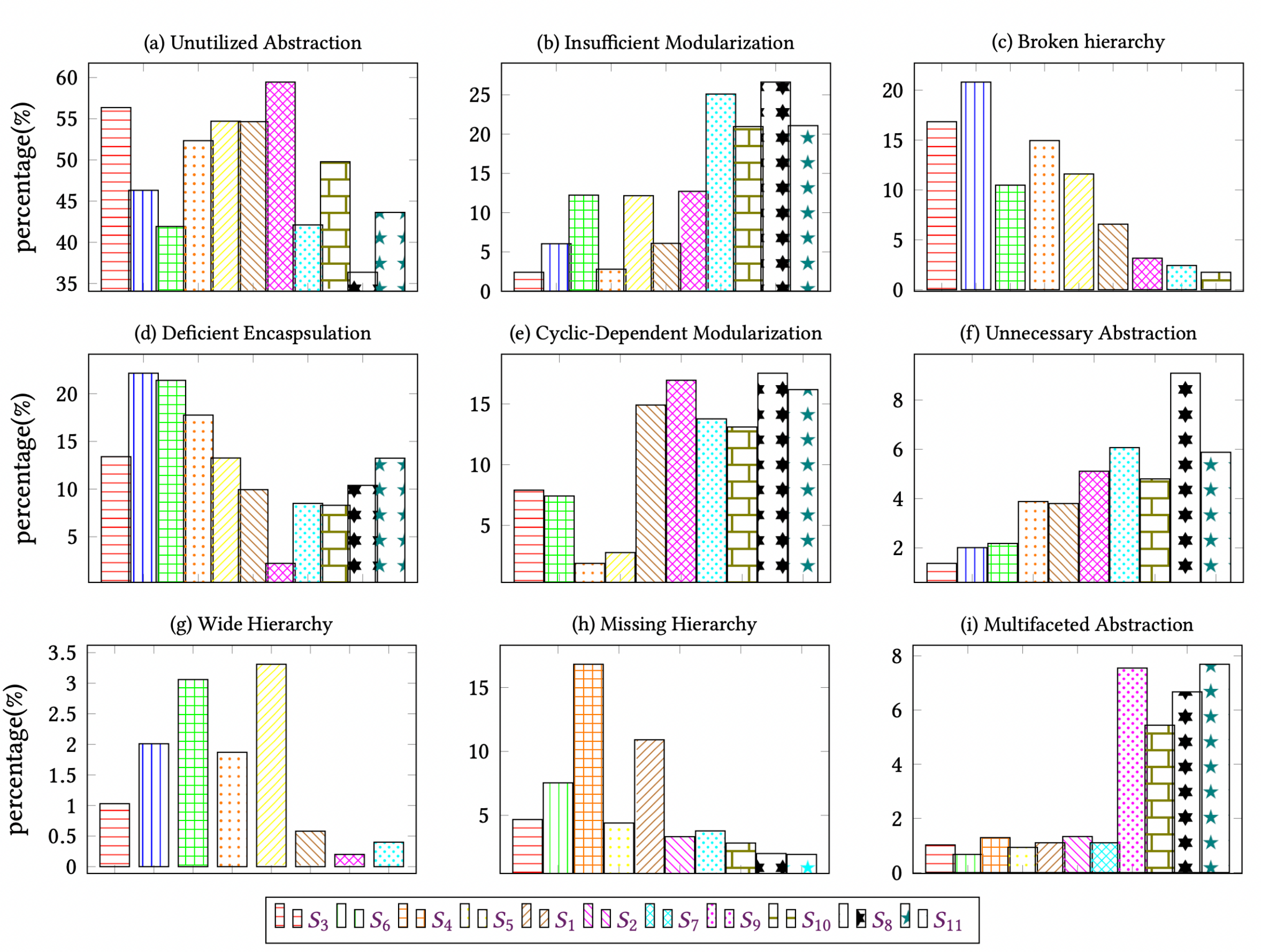}
    \caption{Percentage (\%) of occurrence of specific design smells to the total number of smells}
    \label{PlusPlusCombinedBar}
\end{figure*}

\renewcommand{\arraystretch}{0.90}
\setlength{\tabcolsep}{1pt}
\begin{table*}
\centering
\scalebox{1}{
\begin{tabular}{|p{2.7cm}|p{1cm}|p{1cm}|p{1cm}|p{1cm}|p{1cm}|p{1cm}|p{1cm}|p{1cm}|p{1cm}|}
\hline
 \diaghead{\theadfont DiagColumnmnHead II}%
  {\\~~~Software~}{~Design~~\\ ~Smell~~} & \thead{\rotatebox[origin=c]{90}{\shortstack[4]{\makecell{Unutilized\\ Abstraction}}}} & \thead{\rotatebox[origin=c]{90}{\makecell{Insufficient\\ Modularization}}} & \thead{\rotatebox[origin=c]{90}{\makecell{Broken\\ Hierarchy}}} & \thead{\rotatebox[origin=c]{90}{\makecell{Deficient\\ Encapsulation}}} & \thead{\rotatebox[origin=c]{90}{\makecell{Cyclic\\ Modularization}}} & \thead{\rotatebox[origin=c]{90}{\makecell{Unnecessary\\ Abstraction}}} & \thead{\rotatebox[origin=c]{90}{\makecell{Multifaceted\\ Abstraction}}} & \thead{\rotatebox[origin=c]{90}{\makecell{Wide\\ Hierarchy}}} & \thead{\rotatebox[origin=c]{90}{\makecell{Missing\\ Hierarchy}}} \TBstrut \\ \hline
       \makecell{$S_5$ } & \makecell{ 60 \\ 56\\93.3\%} &  \makecell{4 \\ 3 \\ 75\%} & \makecell{17 \\ 16 \\ 94.1\%} & \makecell{19 \\ 19 \\ 100\%} & \makecell{2 \\ 2 \\ 100\%} & \makecell{8 \\ 8 \\ 100\%} & \makecell{2 \\ 1 \\ 50\%} & \makecell{3 \\ 2 \\ 66.7\%} & \makecell{2 \\ 1 \\ 50\%} \\ \hline
         \makecell{$S_{11}$}  & \makecell{103 \\ 89 \\ 86.4\%} & \makecell{47 \\ 43 \\ 91.5\%} & \makecell{0 \\ 0 \\ 100\%} & \makecell{33 \\ 27 \\ 81.8\%} & \makecell{38 \\ 33 \\ 86.8\%} & \makecell{12 \\ 12 \\ 100\%} & \makecell{18 \\ 17 \\ 94.4\%} & \makecell{0 \\ 0 \\ 100\%} & \makecell{0 \\ 0 \\ 100\%}  \\ \hline
    \end{tabular}
    }
    \caption{Results of the design smell determination mechanism for nine types of smells. \textit{(In each row, the first line identifies the quantity of number of suspected smells detected by the tool, the second line shows the number of actual design smells (true positives), the third line shows the precision as percentage.)} }
\label{fig:precision}
 \end{table*}   
 
\section{Precision and Recall of the Detection Mechanism}
\label{analysis}

In this subsection we analyze the precision and recall for the design smell detection tool. Previous results are capable of detecting four types of design smells for Java \cite{moha2009decor} whereas this approach detects nine types of smells with desirable accuracy. 

The suspicious classes were identified during the detection phase and those were manually analyzed by a team of 16 software professionals to validate the findings. We analyzed the number of detected design smells in Table \ref{fig:top20list3}. Details of the software professionals are highlighted in Table \ref{surveytable}. It should be mentioned that the team of software professionals are from the two companies who designed and developed $S_5$ and $S_{11}$. The classes where design smells were detected by our tool in $S_5$ and $S_{11}$ were presented to the respective software teams who further analyzed it to determine true positives for each smell. We extend the effort to the area of information retrieval and detect precision and recall \cite{frakes1992information}. More specifically, precision identifies the smells out of the total which could be successfully detected. Recall assesses the total number of detected and undetected smells. 


Table \ref{fig:precision} present the precision and recall values of 2 from the list of 11 software analyzed during this research. Overall, precision value for 9 smells were obtained as \textit{81.01\%} for \textit{$S_5$}. For \textit{$S_{11}$}, average precision of \textit{93.43\%} was recorded. The number of suspicious classes which were manually analyzed by six software engineers are 16 and 33 for \textit{$S_5$} and \textit{$S_{11}$} respectively, which are \textit{17.4\%} and \textit{19.0\%} of the total number of classes for the two aforementioned software. The low percentage of classes which were suspected allowed manual analysis of those within a reasonable time frame compared to having to analyze a total of 266 classes otherwise for those two software.

\begin{table}
\begin{tabular}{ |c|c|c|p{3cm}| } 
\hline
\multirow{3}{6em}{Experience} & 4 years and above -> 2  \\
& 2-4 years -> 4\\
& 1-2 years -> 9\\
\hline
\multirow{4}{6em}{Responsibility} & Project manager -> 2  \\
& Software developer -> 8\\
& Quality assurance -> 4\\
& Network administrator -> 8\\
\hline
\multirow{4}{6em}{Demographic} & North America -> 12  \\
& Europe -> 2\\
& Latin America -> 1\\
& Asia -> 1\\
\hline
\multirow{5}{6em}{Java experience} & Spring -> 10  \\
& Hibernate -> 5\\
& GWT -> 4\\
& Grails -> 3\\
& *some respondents have multiple expertise\\
\hline
\end{tabular}
\caption{Demographic and expertise related information about the respondents}
\label{surveytable}
\end{table}

The detection tool has \textit{100\%} recall value since no smell were missed within the scope of the nine types of smells considered in this research. Next the 16 software professionals were assigned a survey to determine the relationship between software development challenges like \textit{"irregular team meetings"} and \textit{"scope creep"} to the occurrence of design smells. 

\section{Survey design for analyzing causes of design smells}
\label{survey}

We conducted a survey among two software organizations out of the eleven considered in this study. The goal of this survey was to find out if the identified design smells were caused by some issues in the software development approach. Earlier we identify the two software a $S_5$ and $S_{11}$ respectively. We try to determine if the cause of the design smells were due to two software engineering malpractices which are \textit{irregular team communication} and \textit{scope creep}. We conduct a survey with the 16 software engineers who designed and developed $S_5$ and $S_{11}$. The same set of 16 respondents were asked to establish the precision and recall of the design smell detection tool.

\subsection{Description of selected projects:} 
\begin{figure}
    \centering
    \includegraphics[width=8cm, height=7cm]{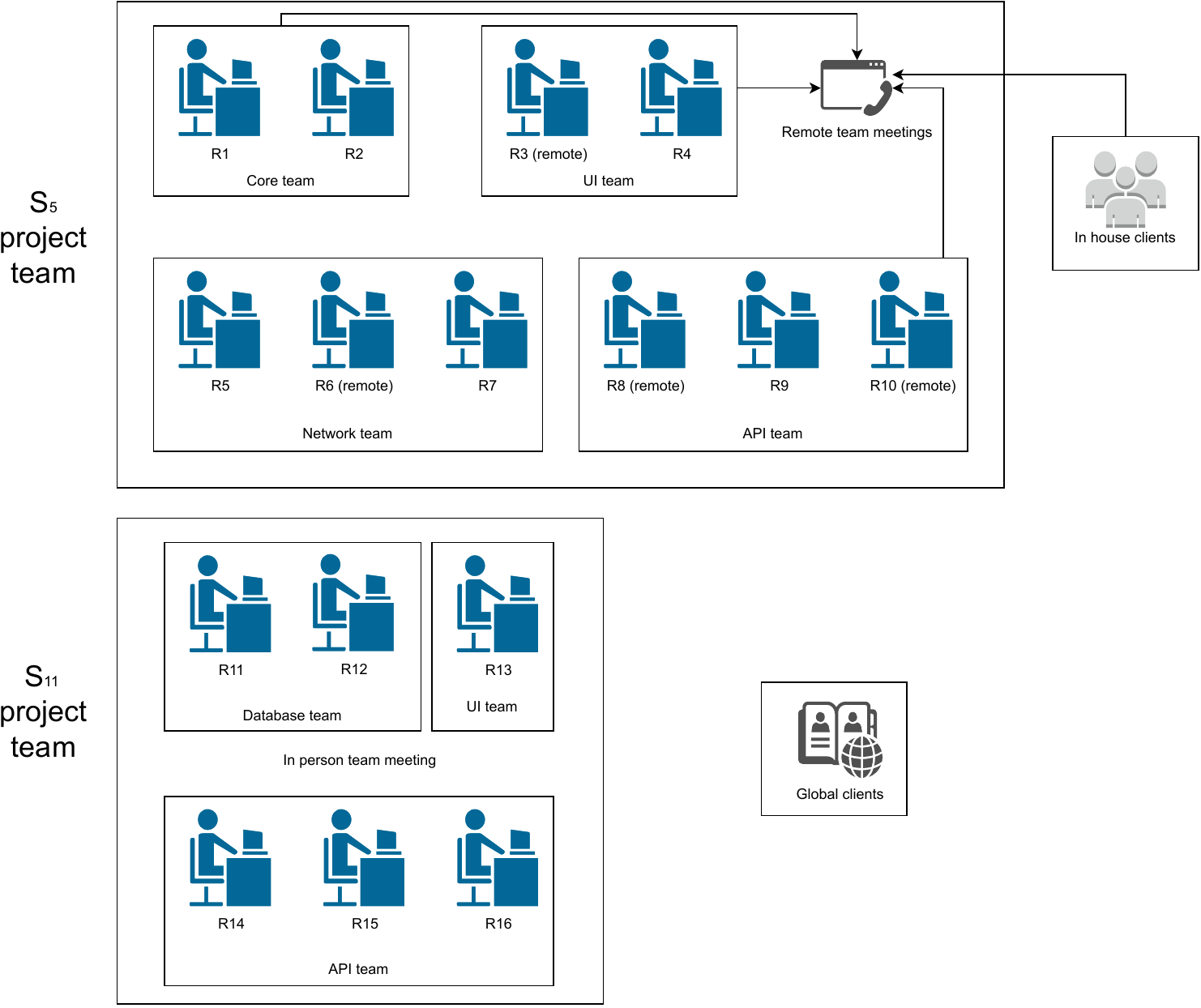}
    \caption{Project team structure for the surveyed software}
    \label{team}
\end{figure}

Figure \ref{team} identifies the teams of respondents who participated in the projects namely $S_5$ and $S_{11}$. Similar to the software names, we preserved confidentiality of the respondents by identifying them as $Rs_1..Rs16$. Project $S_5$ was developed for in-house use. The software was designed and developed by a team of 10 members who are included in the 16 respondents for this survey. The team was divided into subteams among which the core-team was responsible for various activities such as design, and the other subteams were responsible for implementation, and validation of the project. The core-team initially committed to interact with the development teams on a bi-weekly basis. The communication was supposed to be online due to Covid-19 and that some members were located in different geographic locations. The team did not follow any defined software design model to the best of our knowledge.

Project $S_{11}$ was primarily developed for the public and it was made available to be downloaded over the web. The team were able to conduct both in-person and online meetings post-Covid because they were located in the same geographic location. The team aimed to followed incremental software development process, however, members did not conduct regular team communication. 

\subsection{Survey mechanism: }We identified the respondents and selected $S_5$ and $S_{11}$ from our list of 11 software mainly because we were able to contact them via Github and based on our prior association with the two companies who designed and developed these projects. To assure diversity, we focused on these two projects since they had developers from various backgrounds as identified in Table \ref{surveytable}. The respondents primarily came from small-sized companies since we wanted to focus on the software engineering challenges faced by these companies. The interviews were semi-structured and were conducted after the team could give their views on the accuracy and precision of our design smell detection tool. we presented a questionnaire which constituted of 12 questions and required 24.5 minutes to complete on average. We asked whether the design smells which they confirmed as successfully detected by our tool were caused by two software engineering malpractices namely "irregular team communication" or "scope creep". Our goal was to determine if solving these software engineering practices at an early stage of software development significantly impacts the occurrence of design smells.

We conducted this interview separately from the precision calculation activity and identified the mapping between occurrence of the design smell to that of the issues in software engineering practice namely \textit{irregular team communication} and \textit{scope creep}. The questionnaire precisely asked the respondents to relate the cause of the detected design smells to one of the two specific flaws of software design considered here. We recorded all the responses and mapped those to the two software development malpractices. We discuss the survey results in the following.

\subsection{Survey results}
\subsubsection{Irregular team communication}: The respondents related 67.60\% of design smells detected by our tool to the deficiency in regular team meetings. Respondents from both companies stated that team communications were ineffectively managed, and important team discussions were not documented. One responded from $S_5$ stated that \textit{"although we were supposed to have pre-scheduled team meetings, the key stakeholders cancelled team meetings regularly citing conflicts with another high priority meeting, I believe this lack of regularly monitoring progress caused so many design smells in the end"}. Another response from the same company was such that \textit{"we had to miss some team meetings since we were handling multiple projects at one time and hence our time was divided"}. 

Regarding project $S_{11}$ a response to one survey question noted \textit{"at the start, we were asked to implement the software by reading white-papers and case studies of similar projects from internet, however, no meeting was held by the manager highlighting technical aspects, that contributed to the poor design"}. It was seen that both the organizations lacked effective and regular meetings for these two projects. Both the organizations were understaffed at the core team level and regular team meetings could not happen mainly due to the fact that the core team was managing other projects at the same time. Also, it was expected that key requirements of the design would be kept in mind by the developers hence no design documents were generated.

\subsubsection{Scope creep}: Another critical problem which caused significant number of design smells was \textit{scope creep}. It was a phenomenon that caused substantial changes to the project scope and design during development without proper change management procedure. This caused unexpected delays and expensive change incorporation. Scope creep was highlighted as a major challenge for both the teams and they attributed that sudden introduction of new requirements made the design unmanageable and introduced design smells. Following responses obtained from the survey regarding project $S_{5}$ which highlighted this issue. One respondent from the $S_5$ project mentioned \textit{"Our team was initially formed to implement a data transfer system, so it was composed of members who had expertise in networking and programming in Java. However, as days passed and we were finishing the third prototype, the management of our company wanted the software to be able to optimize data transfer via machine learning. To be honest, none of our team members had knowledge on advanced ML techniques so the design has flaws as detected by the tool."} 

Another notable comment we received during survey from a member of $S{11}'s$ project team was that \textit{"change to scope was inevitable, we received a new requirement which drastically changed one method and added many functionalities, it became a god method, we added many for-loops in the method for the various activities it conducted, so the tool found many cyclic dependent modularization design smells in that method. We did not expect the scope of that method to change so much, the new features introduced suddenly also took a lot of time to implement"}. Based on the above response, we highlighted that both the projects suffered from \textit{scope creep} which introduced longer time to market and the final version had design smells. As a result, within the scope of the research we related the cause of the design smells to the two software engineering malpractices.

The outcome of the survey indicates that improved and regular team meetings are required during the design phase to prevent design smells from occurring in the final version of the software. Regular team meetings contribute to a cohesive team and help to clarify design-related questions. Also, key minutes of the team meetings need to be recorded. On the other hand, it has been found that scope creep plays an important role in introducing design smells. Sudden changes in project scope will introduce new requirements, and it is seen within the scope of the two projects analyzed here that engineers find it a challenge to comprehend the design changes in a short time. As a result, project managers need to be aware of not allowing scope creep in their projects and document a change management plan in this regard.

\section{Threats to validity and credibility}
\label{threat}
We identify the common threats to validity as experienced by similar qualitative research. Generalization of the survey results beyond the scope of the sampled respondents needs to be conducted with care. We interviewed only industry experts and not the clients of the software, also we interviewed the team related to the software we tested for design smells, the opinion may vary among teams working on different software projects in the same companies. We limited our scope to small software companies only, however, the findings may not be reflected across different sized companies at different geographic locations.

\section{Related Work}
\label{relatedwork}
A review of exiting tools to detect architectural smells is provided by Azadi et.al. \cite{azadi2019architectural} which groups those based on the detected smells. The authors evaluated 9 tools which are currently in place and ignored those which became obsolete. Smell definitions on which the tools detect the smells are provided and a high level detection mechanism is described by analyzing the current literature. Although the tools which detect architectural smells are provided, accuracy of the tool to aide developers regarding which smells to focus on during different stages of software life cycle was not addressed.

Model depicting the collected efforts required to detect architectural design smells is proposed based on study of related literature \cite{besker2016systematic}. The model identifies negative affects, challenges, refactoring areas and relationship between each activity. The design of the model was motivated from the literature study conducted earlier in the paper. Although the model considers important aspects of design smells, it is a high level design and performance analysis of it is not provided as it was not tested on any software. Also the ability of the proposed architecture to detect smells of software was not discussed.


The affect of developer's seniority, frequency of commits and interval of commits on reducing design debts in a software was evaluated by Alfayez et al.  \cite{Alfayez:2018:ESI:3194164.3194165}. The authors determined that seniority and frequency of commits are negatively correlated to reducing design debt, whereas interval of commits is positively correlated. The authors used multiple statistical analysis tests to validate the affects of developer behavior on the design smells. 

Nahar et al. \cite{10.1145/3510003.3510209} studied the collaboration challenges between ML team and other teams in software projects when the software requires incorporation of an ML component. In this regard, a survey was conducted among software practitioners and the results emphasized on the importance of effective communication and team meeting for successful ML incorporation. However, the impact of \textit{scope creep} on such collaboration have not been studied to the best of our knowledge. 

Although many software analysis tools provide information on various metrics, there is a limitation in the number of open source tools which can detect design smells \cite{biaggi2018architectural}. \textit{AI Reviewer} identifies code and design smells for C++ projects \cite{azadi2019architectural}. \textit{Hotspot Detector} is capable of detecting only 4 design smells in Java \cite{mo2015hotspot}. \textit{Designite} is a commercial software which detects 25 design smells for C\# projects only \cite{7809461}. \textit{Lattix} is another commercial tool which uses dependency matrix to detect modularity violations \cite{wong2011detecting}. Other commercial tools such as \textit{Structure101, Sonargraph} and \textit{Cast} detect cyclic dependency smells \cite{roveda2018identifying}. To the best of our knowledge, although there are tools to detect code smells, those which detect design smells for Java are limited to detection of 4 or less smells.  

Imran et al. \cite{Imran2021} identified the impact of human factors on software sustainability. The authors conducted a survey to show the challenges which software engineers face when incorporating sustainability into their software. They identified human factors which highly impact software sustainability such as \textit{leadership, communication, peer pressure,} and \textit{acknowledgement of efforts}. However, the impact of \textit{irregular team meetings} was not considered. Additionally, how the human factors impact the design of the software was not studied.

\section{Conclusions and Future Work}
\label{conclusion}
This paper detected design smells using an \textit{Abstract Syntax Tree} ($AST$) based tool. Next, it identified specific design qualities and set up relationships between violation of those and occurrence of design smells. After testing the tool on the large volume of LoC and class files of the Java projects analyzed here, the following conclusions were drawn.

\begin{itemize}
    \item The tool was required for correct and time-friendly detection of design smells.
    \item Software engineers could focus on the challenges they faced in terms of team communication and change management of scope to prevent design smells from occurring.
\end{itemize}


In the future, it may also be helpful to perform a longitudinal study that detects how the occurrence of design smells in a program suite is affected by other software design challenges not identified here. A minor but still interesting point is that the pseudo-model tool can be extended to include a new layer showing how design smells change as the software matures.

\begin{acks}
This project is in part sponsored by the National Science Foundation (NSF) under award numbers OAC-1724898, OAC-1842054 and CCF-2007829.
\end{acks}

\bibliography{esse2}
\end{document}